\documentclass[%
 reprint,
superscriptaddress,
 amsmath,amssymb,
 aps,
prb,
]{revtex4-2}

\usepackage{textcomp}
\usepackage{graphicx}
\usepackage{dcolumn}
\usepackage{bm}
\usepackage{upgreek}
\graphicspath{{images/}}
\usepackage{siunitx}  
\usepackage{comment}

\begin{document}

\preprint{APS/123-QED}

\title{
Nonreciprocal magnetoacoustic waves in synthetic antiferromagnets with Dzyaloshinskii--Moriya interaction
}

\author{M. K\"u\ss{}}
 \email{matthias.kuess@physik.uni-augsburg.de.}
 \affiliation{Experimental Physics I, Institute of Physics, University of Augsburg, 86135 Augsburg, Germany\looseness=-1}

\author{M. Hassan}
 \affiliation{Experimental Physics IV, Institute of Physics, University of Augsburg, 86135 Augsburg, Germany\looseness=-1}
 
\author{Y. Kunz}
 \affiliation{Fachbereich Physik and Landesforschungszentrum OPTIMAS, Technische Universit{\"a}t Kaiserslautern, 67663 Kaiserslautern, Germany}

\author{A. H{\"o}rner}
 \affiliation{Experimental Physics I, Institute of Physics, University of Augsburg, 86135 Augsburg, Germany\looseness=-1}

\author{M. Weiler}
 \affiliation{Fachbereich Physik and Landesforschungszentrum OPTIMAS, Technische Universit{\"a}t Kaiserslautern, 67663 Kaiserslautern, Germany}
 
\author{M. Albrecht}
 \affiliation{Experimental Physics IV, Institute of Physics, University of Augsburg, 86135 Augsburg, Germany\looseness=-1}


\date{\today}


\begin{abstract}
The interaction between surface acoustic waves (SAWs) and spin waves (SWs) in a piezoelectric/magnetic thin film heterostructure yields potential for the realization of novel microwave devices and applications in magnonics.
In the present work, we investigate the SAW-SW interaction in a Pt/Co(\num{2})/Ru(\num{0.85})/Co(\num{4})/Pt synthetic antiferromagnet (SAF) composed of two ferromagnetic layers with different thicknesses separated by a thin nonmagnetic Ru spacer layer.  
Because of the combined presence of interfacial Dzyaloshinskii--Moriya interaction (iDMI) and interlayer dipolar coupling fields, the optical SW mode shows a large nondegenerate dispersion relation for oppositely propagating SWs. 
Due to SAW-SW interaction, we observe nonreciprocal SAW transmission in the piezoelectric/SAF hybrid device.
The equilibrium magnetization directions of both Co layers are manipulated by an external magnetic field to set a ferromagnetic, canted, or antiferromagnetic configuration. This has a strong impact on the SW dispersion, its nonreciprocity, and SAW-SW interaction.
The experimental results are in agreement with a phenomenological SAW-SW interaction model, which considers the interlayer exchange coupling, iDMI, and interlayer dipolar coupling fields of the SWs.

\end{abstract}

\maketitle

\section{Introduction}

Surface acoustic waves (SAWs) are widely employed in research and technology nowadays. Efficient excitation and detection of SAWs is possible with metallic grating structures - so-called interdigital transducer structures (IDTs) - on piezoelectric crystals~\cite{R.M.White.1965}. 
These acoustic devices are employed in telecommunications as rf-filters~\cite{Campbell.1998, Morgan.2007} and have further applications, e.g., in microfluidic lab-on-a-chip devices~\cite{Guttenberg.2005, Ding.2013}, quantum acoustics~\cite{Wixforth.1986, Barnes.2000}, and sensors~\cite{Paschke.2017, Lange.2008, Joshi.1994}. 
In general, SAWs show large propagation lengths but are generally reciprocal and offer little control parameters.
%

\begin{figure}
\includegraphics[scale=1]{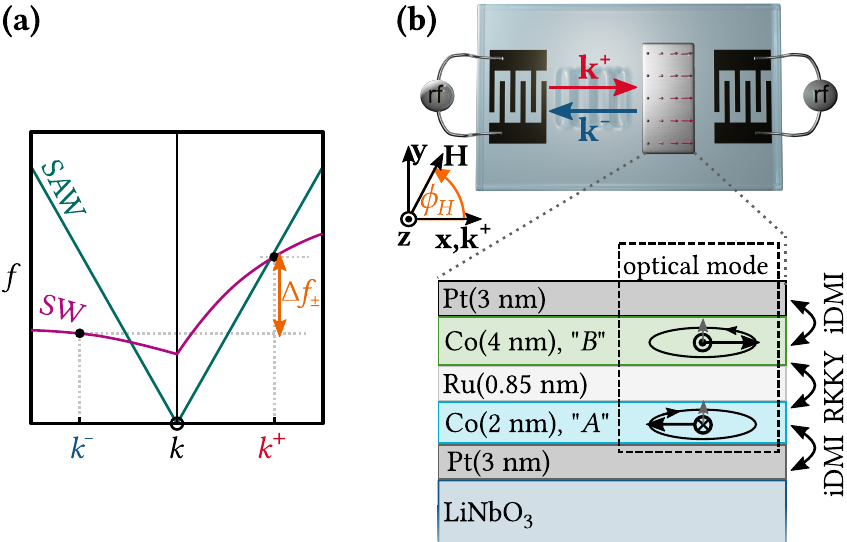}
\caption{
(a)
Interlayer dipolar coupling and iDMI can give rise to a very high nonreciprocity $\Delta f_\pm=f(k^+)-f(k^-)$ of the SW dispersion in a synthetic antiferromagnet (SAF) with iDMI. Thus, the SAW-SW resonance differs for counter-propagating SAWs. 
(b)
Illustration of the investigated magnetoacoustic hybrid device and the coordinate system used. The nonreciprocity of the magnetoacoustically excited SWs in the Pt/Co(2)/Ru/Co(4)/Pt SAF is characterized by different transmission magnitudes of oppositely propagating waves $\mathbf{k}^+$ and $\mathbf{k}^-$. 
For moderate excitation frequencies ($f<\SI{7}{GHz}$) solely the optical mode can be excited in the antiferromagnetic (AFM) configuration. This is schematically shown by the trajectory of the precessing magnetic moments (black arrows) in the Damon-Eshbach geometry ($\phi_0^l=\SI{\pm90}{\degree}$).
\label{fig:1}
}
\end{figure}

In contrast, spin waves (SWs) in magnetic media have low propagation lengths but can be controlled in many ways~\cite{Barman.2021}. 
For instance, the dispersion $f(k)$ of SWs can be reprogrammed by external magnetic fields or electrical currents~\cite{R.A.Gallardo.2019, Ishibashi.2020}. Furthermore, $f(k)$ can show different resonance frequencies for counter-propagating SWs with wave vectors $k^+$ and $k^-$, as depicted in Fig.~\ref{fig:1}(a). This frequency nonreciprocity $\Delta f_\pm = f(k^+)-f(k^-)$ can be pronounced in heavy metal/ferromagnetic films due to the interfacial Dzyaloshinskii--Moriya interaction (iDMI)~\cite{Moon.2013, CortesOrtuno.2013, Nembach.2015, Di.2015, Belmeguenai.2015} and in magnetic bi- and multilayers with nonmagnetic spacer layers via the interlayer dipolar coupling (IDC) fields of the SWs~\cite{Camley.1982, Grunberg.1985, R.A.Gallardo.2019}. 
%
The physical origin of the nonreciprocity lies in the opposite chiralities of counter-propagating SWs. The energy degeneracy is lifted due to chirality-dependent interactions such as iDMI or dipolar interactions.
%
The nonmagnetic spacer layer in magnetic bilayers can give rise to Ruderman–Kittel–Kasuya–Yosida (RKKY)-type~\cite{Bruno.1991} indirect interlayer exchange coupling, which can result in a preferred antiferromagnetic alignment of the equilibrium magnetizations of both ferromagnetic layers. For simplicity, we refer to these layers as synthetic antiferromagnets (SAFs)~\cite{R.A.Duine.2018}. The nonreciprocity $\Delta f_\pm$ in such SAF layers can be especially strong, reconfigurable, and tuneable by the composition of the materials~\cite{R.A.Gallardo.2019}. 
It was recently predicted that the nonreciprocity in SAFs can be even further enhanced by the additional presence of iDMI~\cite{Franco.2020}.

The advantageous properties of SAWs and SWs can be combined in piezoelectric-ferromagnetic hybrid structures~\cite{Kuess.2022-Frontiers}, as shown in Fig.~\ref{fig:1}(b). 
Phonon-magnon coupling is mediated by, e.g., magnetoelastic interaction and opens up the route toward energy-efficient SW excitation and manipulation in the field of magnonics~\cite{Yang.2021}. 
But also the properties of the SAW are greatly affected by resonant magnetoacoustic interaction. 
It was recently demonstrated that resonant magnetoacoustic coupling between SAWs and nonreciprocal SWs can give rise to large nonreciprocal SAW transmission, which may be useful for new devices such as microwave acoustic isolators or circulators~\cite{Verba.2018, R.Verba.2019, Verba.2021, Shah.2020, Ku.2021b}. Hereby, the nonreciprocity of SWs is caused by either iDMI~\cite{Verba.2018, Ku.2020, Xu.2020} or IDC in magnetic bilayers with~\cite{R.Verba.2019, Verba.2021, Shah.2020, Matsumoto.2022} and without~\cite{Ku.2021b} interlayer exchange coupling through nonmagnetic spacer layers. 
%


Here, we experimentally study the impact of a large nonreciprocal SW dispersion $\Delta f_\pm$ of a SAF with iDMI \textit{and} IDC on the SAW transmission in a piezoelectric/magnetic heterostructure, as shown in Fig.~\ref{fig:1}. 
The nonreciprocity $\Delta f_\pm$ of the low-frequency optical SW mode is enhanced by the individual additive nonreciprocal contributions of iDMI and IDC in the investigated Pt/Co(\num{2})/Ru(\num{0.85})/Co(\num{4})/Pt SAF thin film sample (all thicknesses are given in \si{nm}).
%
%
Moreover, the equilibrium magnetization directions of both magnetic Co layers are tuned by an external magnetic field to a ferromagnetic, canted, and antiferromagnetic configuration. 
We demonstrate that these configurations govern the SW resonance frequency, magnetoacoustic driving fields, and SAW transmission.
The experimental results are well reproduced for multiple excitation frequencies by an extended phenomenological model, which considers the interlayer exchange coupling, IDC fields of the SWs, and iDMI.

\section{Theory}



We describe the SAW-SW interaction in SAFs consisting of a bottom magnetic layer $A$  (\SI{2}{nm} Co), a non-magnetic spacer layer, and a top magnetic layer $B$  (\SI{4}{nm} Co), with a phenomenological model.
The thicknesses and saturation magnetizations of the magnetic layers are $d^l$ and $M_s^l$, respectively, with $l=A,B$.
As depicted in Figs.~\ref{fig:1}(b) and \ref{fig:2}(b), the $x$ axis of the $xyz$ coordinate system is parallel to the SAW propagation direction $k^+$ and the $z$ axis is normal to the plane of the magnetic film. The wave vectors $k$ of SAW and SW are defined as $\mathbf{k}=k \hat{\mathbf{x}}$. For counter-propagating waves with wave vectors $|k|$, we write $k^+$ ($k>0$) and  $k^-$ ($k<0$).
In contrast to previous studies on magnetoacoustic interaction in magnetic bilayers~\cite{R.Verba.2019, Shah.2020, Ku.2021b}, we consider SAFs with iDMI, IDC, and bilinear interlayer exchange coupling (IEC). The combination of all these interaction mechanisms gives rise to different configurations of the equilibrium magnetization and a large nonreciprocity $\Delta f_\pm$ of the SW dispersion.

\subsection{Equilibrium state of the magnetization}
\label{sec:Equilibrium M}



First, we calculate the equilibrium orientations $\phi_0^l$ of the magnetizations $\mathbf{M}^l$ in the $xyz$ coordinate system by numeric local energy minimization utilizing a macrospin model~\cite{Bloemen.1994, Zhang.1994, Demokritov.1998, Strijkers.2000, Franco.2020}. 
For both layers, we take into account the Zeeman energy with the external magnetic field magnitude $H$ and direction $\phi_H$ and phenomenological in-plane uniaxial anisotropies with the magnitudes $H_\text{ani}^l$ and the directions $\phi_\text{ani}^l$ with respect to the $x$ axis. 
Thin film shape anisotropy fields are partly compensated by surface anisotropy fields $H_k \hat{\mathbf{z}}$. The equilibrium magnetizations are aligned in the $xy$ plane ($M_z^l=0$). 
The bilinear interlayer exchange coupling with the effective interlayer coupling constant $J_\text{eff}$ favors an antiparallel alignment ($J_\text{eff}<0$) of $\mathbf{M}^A$ and $\mathbf{M}^B$. 
We use Eq.~\eqref{appendix:eq:equilibrium M} of Appendix~\ref{appendix:sec:equilibrium M} to calculate $\phi_0^l$.

\begin{figure}
\includegraphics[scale=1]{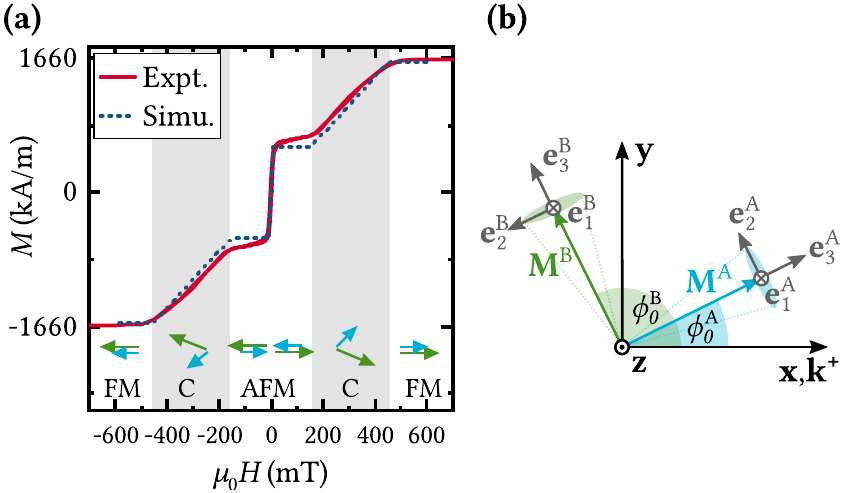}
\caption{
(a)
The $M$-$H$ hysteresis loop of the Pt/Co(2)/Ru/Co(4)/Pt SAF with the magnetic field applied in the $xy$ easy magnetization plane along $\phi_H=\SI{0}{\degree}$.
The magnetization hysteresis loops of experiment and simulation agree well. Both show a FM, C, and AFM configuration, which are additionally schematically depicted by the blue (layer $A$) and green (layer $B$) magnetization vectors. The parameters of the simulation are given in Table~\ref{tab:1} and $\phi_H=\SI{0}{\degree}$ is assumed. 
(b)
The magnetization dynamics and the effective magnetoacoustic driving fields are described in individual ($\mathbf{e}_1^l,\mathbf{e}_2^l,\mathbf{e}_3^l$)-coordinate systems of both layers. Here, the $\mathbf{e}_3^l$-axes are defined by the equilibrium orientations $\phi_0^l$ of the magnetizations $\mathbf{M}^l$.
\label{fig:2}
}
\end{figure}

The calculated equilibrium magnetization orientations $\phi_0^l(\mu_0 H)$ and resulting $M$-$H$ magnetization hysteresis loop of the sample are shown in Figs.~\ref{fig:2}(a) and \ref{fig:3}(c) as a function of the external magnetic field magnitude $\mu_0 H$. The parameters used are listed in Table~\ref{tab:1}. 
The magnetization configuration shows three different magnetic states - namely the ferromagnetic (FM, $\phi_0^A=\phi_0^B$), the canted (C, $\phi_0^A \neq \{\phi_0^B, \phi_0^B \pm \SI{180}{\degree}\}$), and the antiferromagnetic (AFM, $\phi_0^A = \phi_0^B \pm \SI{180}{\degree}$) configuration. While the FM configuration is favored by the Zeeman energy at high fields $|\mu_0 H| \gtrsim \SI{460}{mT}$, the AFM configuration is mediated by the antiferromagnetic interlayer exchange coupling at low fields $|\mu_0 H| \lesssim \SI{150}{mT}$.
The three different magnetization configurations are also experimentally observed in the $M$-$H$ magnetometry measurements in Fig.~\ref{fig:2}(a), which were obtained with the external magnetic field aligned in the plane of a reference sample SAF along $\phi_H=\SI{0}{\degree}$. 
More details about the experimental results will be given later.
The equilibrium magnetization orientations have a strong impact on the SW dispersion and the magnetoacoustic driving fields, as will be discussed next. 

\subsection{Spin wave dispersion} 

The dispersion of a SAW $\omega = c_\text{SAW} |k|$ in a homogeneous substrate with the angular frequency $\omega = 2 \pi f$ is defined by a constant propagation velocity of about $c_\text{SAW} \approx \SI{3500}{m/s}$~\cite{Morgan.2007}, as schematically depicted in Fig.~\ref{fig:1}(a).
In contrast, the dispersion of SWs in SAFs depends on multiple material parameters and its behavior is more complicated.

To calculate the SW dispersion, we follow the approach of Ref.~\cite{Ku.2021b} and solve the coupled linearized Landau--Lifshitz--Gilbert (LLG) equations of both magnetic layers.
To this end, we employ the ($\mathbf{e}^l_1$, $\mathbf{e}^l_2$, $\mathbf{e}^l_3$) coordinate systems, shown in Fig.~\ref{fig:2}(b). Hereby, the $\mathbf{e}^l_3$-directions correspond to the equilibrium magnetization orientations, the $\mathbf{e}^l_2$-directions are aligned in the plane of the magnetic film, and the $\mathbf{e}^l_1$-directions are parallel to the film normal.
%
The magnetization precession is defined by the transverse magnetization components $M^l_{1,2} = M^l_s m^l_{1,2}$ and is assumed to be small ($M^l_{1,2} \ll M^l_s$).

We formulate the effective fields of the LLG equations in the  ($\mathbf{e}^l_1$, $\mathbf{e}^l_2$, $\mathbf{e}^l_3$) coordinate systems as
\begin{equation}
    \mathbf{H}_\text{eff,123}^l = \mathbf{H}_\text{eff,123,intra}^l + \mathbf{H}_\text{eff,123,IEC}^l + \mathbf{H}_\text{eff,123,IDC}^l +
    \begin{pmatrix}
        h_1^l \\ h_2^l \\ 0
    \end{pmatrix}.
\end{equation}
with the intralayer effective fields $ \mathbf{H}_\text{eff,123,intra}^l$~\cite{Ku.2020} caused by the (i) Zeeman energy, (ii) in-plane uniaxial magnetic anisotropy, (iii) out-of-plane surface anisotropy, (iv) intralayer exchange interaction, (v) intralayer dipolar fields, and (vi) iDMI fields $ {\bf H}_\text{eff,123,DMI}^l$.
The interlayer coupling is mediated by bilinear exchange fields $\mathbf{H}_\text{eff,123,IEC}^A$ and the dipolar fields $\mathbf{H}_\text{eff,123,IDC}^l$ of the SWs.
The terms for $ {\mathbf H}_\text{eff,123,DMI}^l$, $\mathbf{H}_\text{eff,123,IEC}^l$ and $\mathbf{H}_\text{eff,123,IDC}^l$ are taken from Refs.~\cite{Moon.2013, R.A.Gallardo.2019}, rewritten in the ($\mathbf{e}^l_1$, $\mathbf{e}^l_2$, $\mathbf{e}^l_3$) coordinate systems, and given in Eqs.~\eqref{eq:HeffDMI}-\eqref{eq:HeffIDC} in Appendix~\ref{appendix:dipolar fields}.
Moreover, the SAW-SW interaction is mediated by small time-varying magnetoacoustic driving fields $h_{1,2}^l$, as defined in the next section.
From the LLG equations, we calculate the $4 \times 4$ magnetic susceptibility tensor which describes the magnetic response
\begin{equation}
    \begin{pmatrix}
     M_1^A \\  M_2^A \\  M_1^B \\  M_2^B
    \end{pmatrix}
    =
    \bar{\chi}
    \begin{pmatrix}
    h_1^A \\  h_2^A \\  h_1^B \\  h_2^B
    \end{pmatrix}.
    \label{eq: def mag. suszeptibility}
\end{equation}
to $h_{1,2}^l$.
The SW resonance frequency $f(\mu_0 H)$ is obtained by setting $\det (\bar{\chi}^{-1}) = 0$ and taking the real part of the solution. The linewidth $\Delta f_\text{HWHM}$ is given by the corresponding imaginary part. 
Here, we neglect mutual spin pumping between both magnetic layers of the SAF~\cite{Heinrich.2003, Sorokin.2020}.

\begin{figure}
\includegraphics[scale=1]{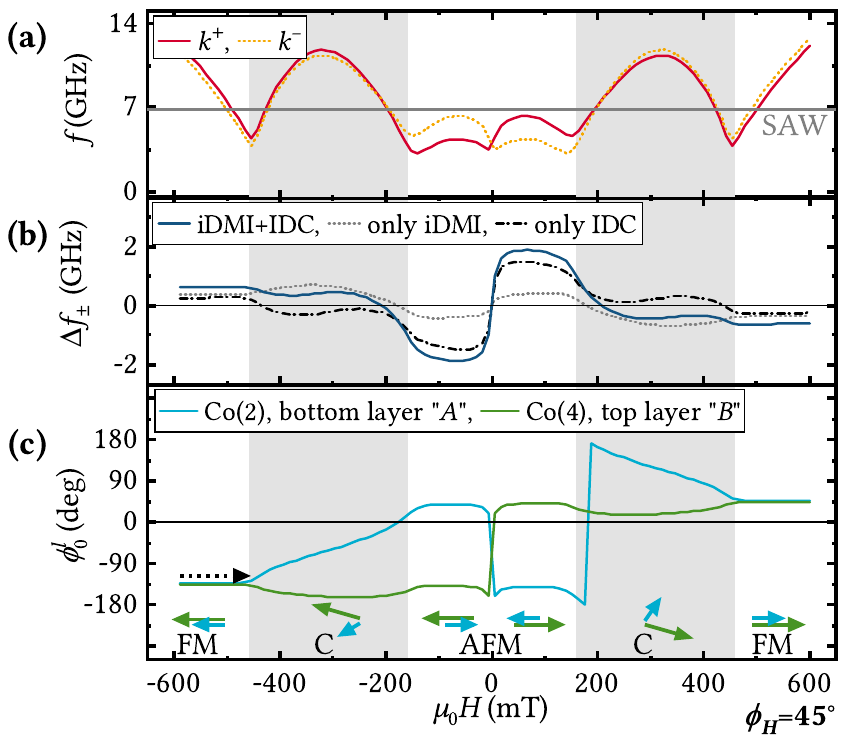}
\caption{
\label{fig:3}
(a)
The calculated resonance frequency of the optical SW mode as a function of the external magnetic field $\mu_0 H$ shows nondegenerated frequencies for counter-propagating SWs with wave vectors $k^+$ and $k^-$. Calculated for $\phi_H = \SI{45}{\degree}$ and $|k^\pm|=\SI{13.1}{\micro m^{-1}}$.
(b)
The large frequency nonreciprocity $\Delta f_\pm = f(k^+)-f(k^-)$ can be understood as an additive interplay of both individual contributions of iDMI and IDC~\cite{Franco.2020}.
(c)
The equilibrium orientations $\phi_0^l$ of the magnetizations $\mathbf{M}^l$, which are schematically depicted by the arrows govern the nonreciprocal SW resonance.
}
\end{figure}

As an example, we calculate the SW resonance $f(\mu_0 H)$ of our Pt/Co/Ru/Co/Pt sample in Fig.~\ref{fig:3}(a). 
The assumed parameters are summarized in Table~\ref{tab:1}, $\phi_H=\SI{45}{\degree}$ is the external magnetic field direction and $|k^\pm|=\SI{13.1}{\micro m^{-1}}$ is the wave vector of a SAW with a frequency of \SI{6.77}{GHz}. 
Only the low-frequency SW mode, which we refer to the SW optical mode~\cite{Franco.2020}, is shown.  The resonance frequency of the high-frequency SW acoustic mode is much higher ($f>\SI{20}{GHz}$) than the SAW excitation frequency ($\SI{6.77}{GHz}$) because of the antiferromagnetic interlayer exchange coupling.
Local minima in the SW resonance frequency appear where the static magnetization configuration (FM, C, AFM) transforms into another one.

In addition, counter-propagating SWs with wave vectors $k^+$ and $k^-$ have shifted resonance frequencies. The origin of the large frequency nonreciprocity $\Delta f_\pm = f(k^+)-f(k^-)$ is characterized in Fig.~\ref{fig:3}(b) in more detail.
Since iDMI and IDC are well-known to cause frequency nonreciprocity, the impact of both individual contributions on $\Delta f_\pm$ is calculated by setting the corresponding other effective field $\mathbf{H}_\text{eff,123,IDC}^l$ or $\mathbf{H}_\text{eff,123,iDMI}^l$ to zero.
A constructive superposition of $\Delta f_\pm$ caused by iDMI and IDC results in a very large frequency nonreciprocity of more than \SI{2}{GHz} in the AFM configuration at $\phi_H=\SI{45}{\degree}$.
The impact of iDMI of the top and bottom ferromagnetic layer on $\Delta f_\pm$ is expected to have additive contributions in the AFM configuration~\cite{Franco.2020}.
The nonreciprocity $\Delta f_\pm$ is low in the FM configuration in comparison to the AFM configuration. Furthermore, the sign of $\Delta f_\pm$ switches between the AFM  and FM configuration. 
A more detailed theoretical discussion about the enhancement of $\Delta f_\pm$ in different antiferromagnetically coupled magnetic bilayers with iDMI and IDC was recently presented by A. Franco and P. Landeros~\cite{Franco.2020}. The Pt/Co(\num{2})/Ru(\num{0.85})/Co(\num{4})/Pt SAF of our study shows a similar behavior as the considered Pt/Co(\num{1.5})/Cu(\num{0.95})/Co(\num{2})/Pt SAF in their work~\cite{Franco.2020}.


\subsection{Magnetoacoustic driving fields} 

The propagating SAW causes strain $\varepsilon_{ij}$ and lattice rotation $\omega_{ij}$ in the SAF with $i,j=x,y,z$. Via SAW-SW coupling mechanisms, such as inverse magnetostriction, the SAW modulates the magnetic free energy. 
The corresponding effective magnetoacoustic driving field as a function of SAW power $P_\text{SAW}$ in the ($\mathbf{e}^l_1$, $\mathbf{e}^l_2$) plane can be written~\cite{Ku.2020} as
\begin{equation}
\begin{pmatrix}
    h_1^l \\ h_2^l
\end{pmatrix} 
=
\begin{pmatrix}
    \tilde{h}_1^l \\
    \tilde{h}_2^l
\end{pmatrix}
\sqrt{ \frac{k^2}{R \, \omega W}} \sqrt{P_\text{SAW}(x)}~\text{e}^{i(kx-\omega t)}.
\label{eq:drivingfields1}
\end{equation}
Here, $W$ is the width of the aperture of the IDT and $R$ is a constant~\cite{Robbins.1977}. 
The normalized effective magnetoelastic driving fields $\tilde{h}_1^l$ and $\tilde{h}_2^l$ of a Rayleigh wave with strain components $\varepsilon_{ij= xx,xz} \neq 0$ are~\cite{Dreher.2012,Ku.2020}
\begin{equation}
    \begin{pmatrix}
        \tilde{h}_1^l \\
        \tilde{h}_2^l
    \end{pmatrix}
 =
 \frac{2}{\mu_0}
 \left[
    b_1^l \tilde{a}_{xx}
    \begin{pmatrix}
           0 \\
        \sin{\phi_0^l} \cos{\phi_0^l}
    \end{pmatrix}
    + b_2^l \tilde{a}_{xz}
        \begin{pmatrix}
            \cos{\phi_0^l} \\
            0
        \end{pmatrix}
 \right],
\label{eq:drivingfields2}
\end{equation}
where $b_{1,2}^l$ are the magnetoelastic coupling constants for cubic symmetry of the ferromagnetic layers~\cite{Dreher.2012,Kittel.1958}, $\tilde{a}_{ij} = \varepsilon_{ij,0} / (|k| |u_{z,0}|)$ are the normalized amplitudes of the strain, and $\varepsilon_{ij,0}$ are the complex amplitudes of the strain. Furthermore, $u_{z,0}$ is the amplitude of the lattice displacement in the $z$ direction. 
Because the wavelength $\lambda=(2 \pi)/|k|$ of the SAW ($\lambda \gtrsim \SI{500}{nm}$) is much larger than the film thickness of the SAF ($\approx \SI{10}{nm}$) we assume constant strain in $z$ direction.
For the sake of simplicity, we neglect non-magnetoelastic interactions, like magneto-rotation coupling~\cite{Maekawa.1976, Xu.2020, Ku.2020}, spin-rotation coupling~\cite{Matsuo.2011, Matsuo.2013, Kobayashi.2017}, and gyromagnetic coupling \cite{Kurimune.2020}.

\begin{figure}
\includegraphics[scale=1]{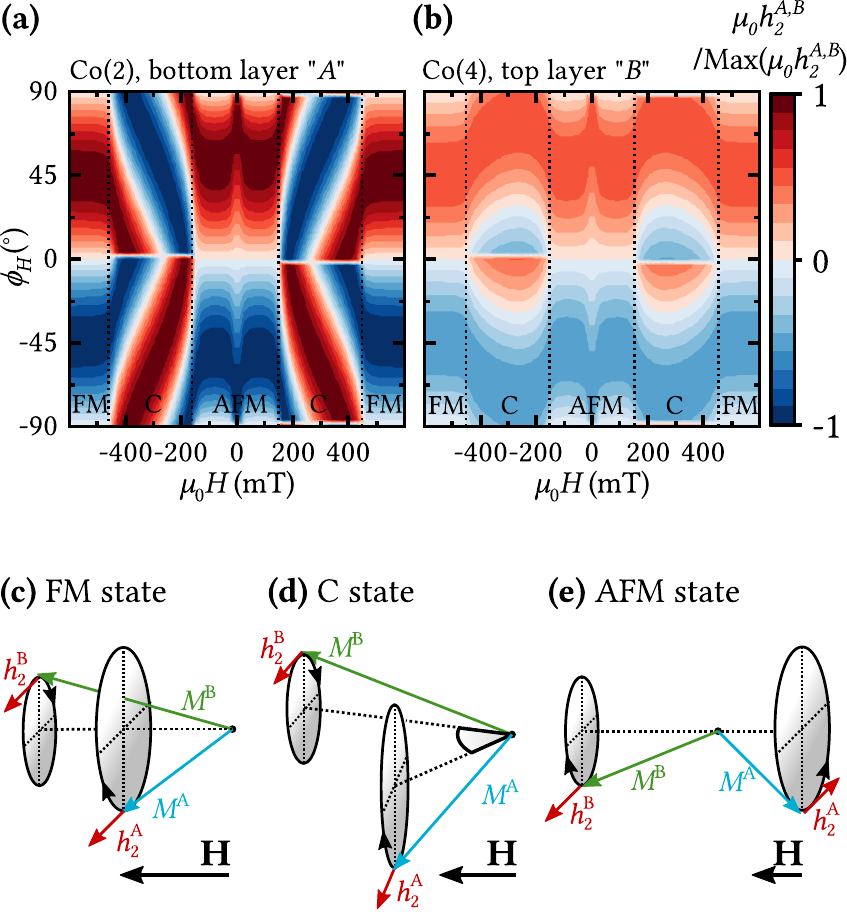}
\caption{
\label{fig:4}
Magnetoacoustic driving field and impact on the excited SW modes in the investigated SAF sample.
(a),(b) 
The dominating in-plane component $h_2^l$ of the effective magnetoacoustic driving field of a Rayleigh-type SAW in the individual magnetic layers of the SAF is calculated with Eq.~\eqref{eq:drivingfields2} on basis of the equilibrium magnetization directions $\phi_0^l$ and the parameters in Table~\ref{tab:1}.
The symmetry of $|h_2^l|$ follows the well-known four-fold symmetry~\cite{Weiler.2011} in the FM and AFM configuration, but is more complicated in the C configuration. 
(c)--(e)
Schematic drawings of the excited optical SW modes and driving fields $h_2^l$, which can potentially be excited in the FM, C, and AFM configuration for $\phi_H = \SI{45}{\degree}$. Here, the acoustic SW modes cannot be magnetoacoustically excited since its resonance frequency is much higher ($f>\SI{20}{GHz}$) than the SAW excitation frequency ($\SI{6.77}{GHz}$).
While the magnetoelastic excitation of SWs in the AFM configuration is efficient, it is not efficient in the FM configuration. 
The drawing of the C configuration depicts only one possible situation. 
The direction of $h_2^l$ is defined by the direction of the $\mathbf{e}_2^l$ axes (see Fig.~\ref{fig:2}). 
}
\end{figure}

The magnetoacoustic driving fields result in a torque on the equilibrium magnetizations of both individual magnetic layers. 
Whether these torques add up or cancel for the excitation of SWs in SAFs depends on the directions of the magnetoacoustic driving fields and on the SW modes which can potentially be excited in the different magnetization configurations (FM, C, AFM). 
To demonstrate the impact of the magnetoacoustic driving fields on the excited SW modes, we calculated from the magnetization equilibrium directions $\phi_0^l$ the magnitude of $h_2^l$ of a purely longitudinal strain wave with $\varepsilon_{xx} \neq 0$ for external magnetic field sweeps from  \SI{-600}{mT} to \SI{600}{mT} at multiple angles $|\phi_H| \leq \SI{90}{\degree}$. The results are shown in Figs.~\ref{fig:4}(a) and \ref{fig:4}(b).

For the FM  and AFM configuration, the magnetoelastic driving field $|h_2^l|$ of both magnetic layers follows the well-known four-fold symmetry~\cite{Weiler.2011} with maxima at $\phi_H=\SI{\pm 45}{\degree}$ and minima at $\phi_H=\SI{0}{\degree}, \SI{\pm 90}{\degree}$. For the C configuration, the symmetry of $|h_2^l|$ is more complicated because the equilibrium magnetization is not parallel or antiparallel to the external magnetic field.
Note that the direction of $h_2^l$ in the fixed $xyz$ coordinate system depends on the direction of the $\mathbf{e}_2^l$-axes [see Fig.~\ref{fig:2}(b)].

In Figs.~\ref{fig:4}(c)--\ref{fig:4}(e) we show an illustration of the driving fields $h_2^l$ and the SW modes, which can potentially be excited in the FM, C, and AFM configuration of the investigated Pt/Co/Ru/Co/Pt SAF for $\phi_H = \SI{45}{\degree}$. 
The optical SW mode in the FM configuration is described by an anti-phase magnetization precession~\cite{Franco.2020}. In contrast, the magnetization precession of the optical SW mode in the AFM configuration is in-phase in the $\mathbf{e}^l_1$ direction and anti-phase in the $\mathbf{e}^l_2$ direction~\cite{R.Verba.2019}.
The drawing of the C configuration depicts only one possible situation and is in general more complicated, since the SW mode and driving field change with the external magnetic field magnitude.
Together with the right-handed magnetization-precession, the magnetoelastic excitation of SWs in the AFM configuration is efficient, whereas it is not efficient in the FM and depicted C configuration.

\subsection{Absorbed power} 

The total absorbed power $P_\text{abs}$ of the SAW, caused by SAW-SW interaction, is the sum of the absorbed power of the individual layers. Following Ref.~\cite{Ku.2021b}, we write
\begin{align}
&P_\text{abs} = P_0
\left(
1 - \text{exp}
\left\{
- \tilde{C} ~\text{Im}
\left[
\begin{pmatrix}
    d^A \tilde{h}_1^A \\ d^A  \tilde{h}_2^A \\ d^B \tilde{h}_1^B \\ d^B \tilde{h}_2^B
\end{pmatrix}^{\mkern -10mu *}
 \bar{\chi}
\begin{pmatrix}
    \tilde{h}_1^A \\\tilde{h}_2^A \\ \tilde{h}_1^B \\ \tilde{h}_2^B
\end{pmatrix}
\right]
\right\}
\right)
\nonumber \\
& \text{with}~ \tilde{C} = \frac{1}{2} \mu_0 l_f \left( \frac{k^2}{R} \right),
\label{eq:PabsDGLsol}
\end{align}
where $l_f$ is the length of the magnetic thin film along the $x$ axis.
Finally, to directly simulate the experimentally determined relative change of the SAW transmission $\Delta S_{ij}$ on the logarithmic scale, we use
\begin{equation}
\Delta S_{ij}
= 10 \lg \left( \frac{P_0 - P_\text{abs}}{P_0} \right) 
\ \text{with} \  
ij = 
\left\{
\begin{array}{ll}
21, & \text{for}~k^+ \\
12, & \text{for}~k^- \\
\end{array}
\right.
\label{eq:5:S21Final}
\end{equation}
for SAWs propagating parallel ($k^+$) and antiparallel ($k^-$) to the $x$ axis.

\section{Experimental Setup}


We used magnetron sputter deposition for the preparation of the Pt(\num{3})/Co(\num{2})/Ru(\num{0.85})/Co(\num{4})/Pt(\num{3})/Si$_3$N$_4$(\num{3}) SAF on a piezoelectric Y-cut Z-propagation LiNbO$_3$ substrate. 
The deposition parameters are detailed in Appendix~\ref{appendix:sample preparation}. 
In such multilayers made out of Co/Ru($d_s$)/Co especially strong antiferromagnetic interlayer exchange coupling has been observed for $\SI{0.5}{nm} \lesssim d_s \lesssim \SI{1}{nm}$~\cite{Ounadjela.1992, Parkin.1990, Bloemen.1994}.
Furthermore, the Pt layers are known to induce strong iDMI in Co~\cite{Yang.2015, Cho., Belmeguenai.2015}.
To obtain a large nonreciprocity $\Delta f_\pm$ of the optical SW mode by a combined action of iDMI and IDC, we follow the design rules of Ref.~\cite{Franco.2020}. 
The bottom Co(2)-layer has a lower natural frequency and a larger thickness-averaged effective DMI constant $D_\text{eff}^A$ than the top Co(4)-layer because of its lower film thickness and lower effective magnetization $M_\text{eff}^l=M_s^l-H_k^l$.
Therefore, the Co(2)-layer governs the low frequency optical SW mode of the SAF which should be strongly affected by the relatively large $D_\text{eff}^A$.

The SAW transmission measurements are carried out with two single finger IDTs that are made out of Ti(5)/Al(70), have three finger pairs with a periodicity of \SI{1.7}{\micro m}, and can be operated at multiple higher harmonic resonance frequencies of up to $\sim\SI{7}{GHz}$. The Y-cut Z-propagation LiNbO$_3$ substrate gives rise to Rayleigh-type SAW excitation~\cite{Morgan.2007}.
More details about the IDT geometry are given in the Supplemental Material of Ref.~\cite{Ku.2020}.

The $M$-$H$ hysteresis loop of the prepared SAF was measured by superconducting quantum interference device-vibrating sample magnetometry (SQUID-VSM) and is presented in Fig.~\ref{fig:2}(a). The SAF shows the expected behavior with FM, C, and AFM magnetization configurations.

On additionally prepared Pt(\num{3})/Co(\num{2})/Ru(\num{0.85})/Si$_3$N$_4$(\num{3}) and Ru(\num{0.85})/Co(\num{4})/Pt(\num{3})/Si$_3$N$_4$(\num{3}) single reference samples SQUID-VSM magnetometry and broadband ferromagnetic-resonance (bbFMR) measurements were carried out to further characterize the magnetic properties of the individual Co layers.

The SAW transmission of our delay line device was measured by a vector network analyzer. Based on the low propagation velocity of the SAW, a time-domain gating technique was employed to exclude spurious signals~\cite{Hiebel.2011}, in particular, electromagnetic crosstalk.
We use the relative change of the background-corrected SAW transmission signal
\begin{equation}
    \Delta S_{ij} (\mu_0 H) = S_{ij} (\mu_0 H) - S_{ij} (\SI{600}{mT})
\end{equation}
to characterize SAW-SW coupling. Here, $\Delta S_{ij}$ is the magnitude of the complex transmission signal with $ij =21,12$. In all measurements, the magnetic field is swept from \SI{-600}{mT} to \SI{+600}{mT} corresponding to full magnetic saturation.

\section{Discussion}
\label{sec:Discussion}

\subsection{Frequency dependence}

\begin{figure*}
\includegraphics[scale=1]{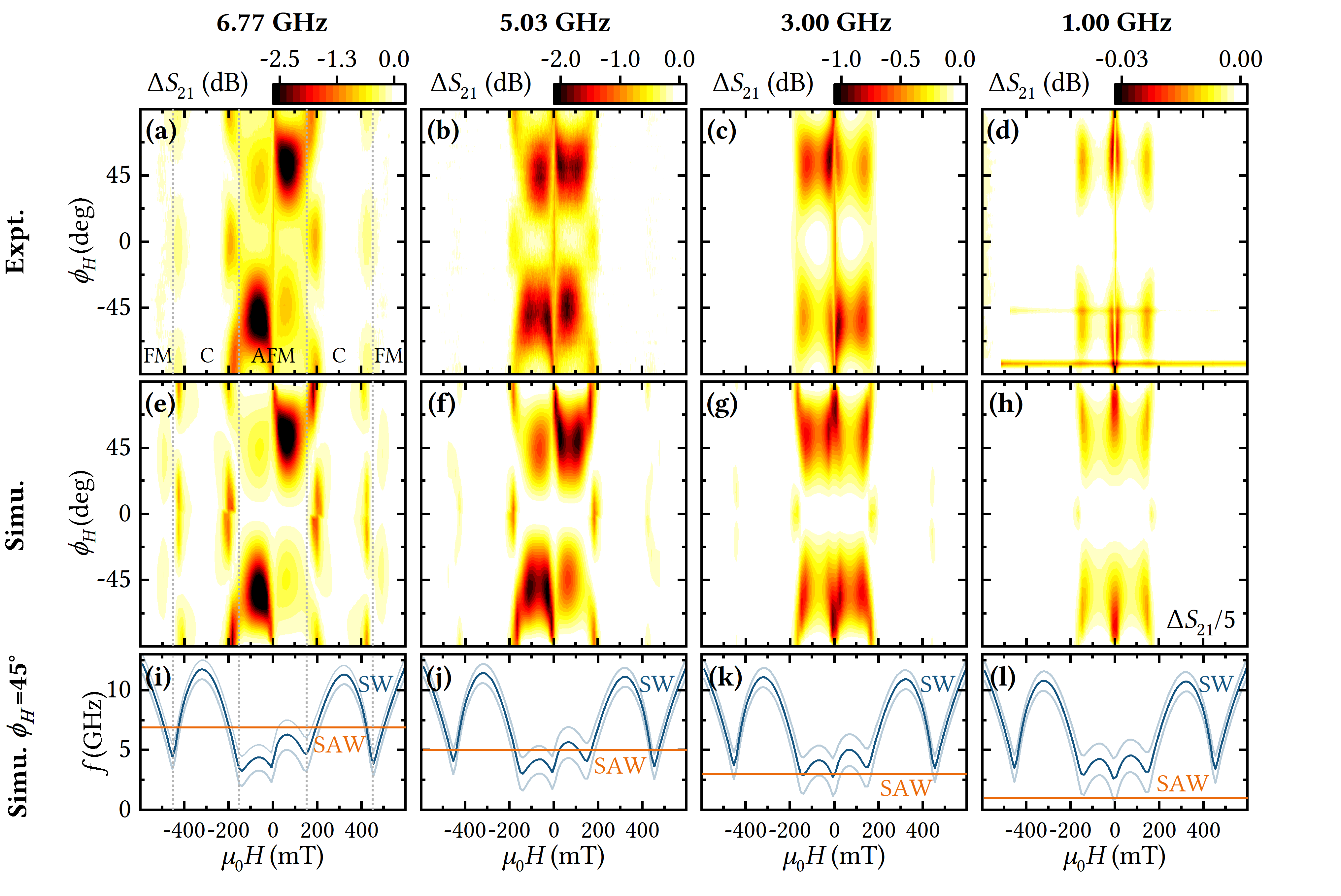}
\caption{
(a)--(h)
The frequency dependence of the magnetoacoustic transmission $\Delta S_{21}(\mu_0 H,\phi_H)$ was determined as a function of external magnetic field magnitude and orientation for all four harmonic resonances of the acoustic delay line.
Experiment (Expt.) and simulation (Simu.) show good agreement. 
(i)--(l) 
The SW resonances $f$ (blue), SW linewidths indicated by $f \pm \Delta f_\text{HWHM}$ (light blue), and SAW excitation frequencies (orange) at $\phi_H = \SI{45}{\degree}$. Both, simulations (e)--(h) and SW resonances (i)--(l) were calculated with the parameters of Table~\ref{tab:1} and the corresponding wave vector of the SAW ($k_\text{SW}=k_\text{SAW}$). 
The nonreciprocity of $\Delta S_{21}(\mu_0 H,\phi_H)$ and of the SW resonance $f(\mu_0 H)$ decrease with decreasing frequency. 
\label{fig:5}
}
\end{figure*}

In Figs.~\ref{fig:5}(a)--\ref{fig:5}(d), we show the magnetoacoustic transmission $\Delta S_{21}$ of the prepared Pt/Co/Ru/Co/Pt SAF as a function of external magnetic field magnitude $\mu_0 H$ and orientation $\phi_H$. 
The measurements were carried out for all four harmonic frequencies $f = \num{1.00}, \num{3.00}, \num{5.03}, \SI{6.77}{GHz}$ and corresponding wave vectors $k = \num{1.9}, \num{5.8}, \num{9.8}, \SI{13.1}{\micro m^{-1}}$ of the acoustic delay line. 
The vertical dotted lines indicate the transitions of the static magnetization configurations (FM, C, AFM) according to the $M$-$H$ hysteresis loop in Fig.~\ref{fig:2}(a). 
In comparison to earlier studies performed on magnetic single layers~\cite{Weiler.2011, Dreher.2012, Gowtham.2015, Ku.2020}, magnetic bilayers with zero interlayer exchange coupling~\cite{Ku.2021b}, and symmetric SAFs with $d^A = d^B$~\cite{Shah.2020, Matsumoto.2022}, many resonances show up for the asymmetric SAF with $d^A \neq d^B$.
The resonances in the FM  and AFM configuration are at maximum at $|\phi_H| \approx \SI{45}{\degree}$. In contrast, the resonances in the C configuration show up at $|\phi_H| \approx \SI{0}{\degree}, \SI{90}{\degree}$.
Whereas the resonances in the FM  and C configuration have a low intensity and relatively narrow linewidth, the intensity and linewidth of the resonances in the AFM configuration are large.

The simulations of $\Delta S_{21}(\mu_0 H, \phi_H)$ are shown in Figs.~\ref{fig:5}(e)--\ref{fig:5}(h).  
Moreover, we simulate the SW resonances $f$, SW linewidths indicated by $f \pm \Delta f_\text{HWHM}$, and SAW excitation frequencies in Figs.~\ref{fig:5}(i)--\ref{fig:5}(l) for $\phi_H=\SI{45}{\degree}$. 
Note that the SW resonances $f(\mu_0 H)$ are in general not constant, but are a function of $\phi_H$. 
All simulations were carried out with the parameters given in Table~\ref{tab:1}. 
We took an increase of $\alpha^l$ with decreasing excitation frequency~\cite{Ku.2020} into account, which is modeled by Eq.~\eqref{eq:alphaEff}. More details about the simulation parameters can be found in Appendix~\ref{appendix:simulation parameters}.

\begin{table*}
\caption{
\label{tab:1}
Parameters to simulate the equilibrium magnetization directions, SW resonance frequencies, magnetoacoustic driving fields, and
the magnetoacoustic transmission $\Delta S_{21}$ in Figs.~\ref{fig:2}--\ref{fig:6} at \SI{6.77}{GHz}. For the simulations at lower frequencies in Fig.~\ref{fig:5}, the effective SW damping constant $\alpha$ is calculated by Eq.~\eqref{eq:alphaEff} with the Gilbert damping ($\alpha_\text{FMR}^A = 0.031$, $\alpha_\text{FMR}^B = 0.019$) and inhomogeneous broadening ($\mu_0 \Delta H_\text{FMR}^A = \SI{17}{mT}$, $\mu_0 \Delta H_\text{FMR}^B = \SI{5}{mT}$) from broadband FMR experiments. For the simulation of $\Delta S_{12}$, the sign of the normalized strain $\tilde{a}_{xz}$ is inverted. Furthermore, the bilinear interlayer exchange coupling constant $J_\text{eff}=\SI{-0.95}{mJ/m^2}$ is extracted from the $M$-$H$ hysteresis loop and we assume $c_\text{SAW}=\SI{3240}{m/s}$~\cite{Note1} for the SAW propagation velocity in the SAF. More details about the simulation parameters are presented in Appendix~\ref{appendix:simulation parameters}.
}
\begin{ruledtabular}
\begin{tabular}{ccccccccccccc}
 & layer $l$ & $d^l$ & $g^l$ & $M_s^l$ & $H_k^l$ & $A_\text{ex}^l$  & $D_\text{eff}^l$ & $\mu_0 H_\text{ani}^l$ & $\phi_\text{ani}^l$ & $\alpha^l$ & $b_1^l \tilde{a}_{xx}$ & $b_2^l \tilde{a}_{xz}$ \\
 &  & (\si{nm})  &   & (\si{kA/m})  & (\si{kA/m})  & (\si{pJ/m})  & (\si{mJ/m^2})  & (\si{mT})  & (\si{\degree})  &   & (\si{T})  & (\si{T}) \\
 \hline
Co(4) & $B$ &  4  & 2.315  & 1630  & 667  & 31~\cite{Bertotti.1998, Franco.2020}  & -0.35  & 4  & 0  & 0.031  &  1.0 & \num{0.22}\text{i} \\
Co(2) & $A$  & 2  &  2.317 & 1580  & 1085 & 31~\cite{Bertotti.1998, Franco.2020}  & +0.70  & 0  & 0  & 0.071  &  1.9  & \num{0.42}\text{i}\\
\end{tabular}
\end{ruledtabular}
\end{table*}

%
%

The simulation and experiment displayed in Fig.~\ref{fig:5} show a rather good quantitative agreement with respect to all salient features such as resonance fields, linewidth, and transmission magnitude in the FM, C, and AFM configuration.
Only the simulation of $\Delta S_{21}$ at \SI{1}{GHz} shows about five times too large amplitudes in $\Delta S_{21}$. 
The magnetoacoustic resonances at \SI{1}{GHz} in Figs.~\ref{fig:5}(d) and \ref{fig:5}(h) result from highly off-resonant excitation and occur where the static magnetization transforms from AFM  to C configuration and around zero field switching. 
The too-large amplitudes of the simulation might be caused by deviations of the magnetization reversal from the macrospin approach (domain formation, etc.), which could be especially significant where the static configuration (FM, C, AFM) transforms into another one and around zero field. 


In the AFM configuration at $\phi_H > 0$ and $0 < \mu_0 H < \SI{150}{mT}$, the resonance with high linewidth at \SI{6.77}{GHz} changes to two resonances at \SI{5.03}{GHz}, as shown in Figs.~\ref{fig:5}(a) and \ref{fig:5}(b). If $f$ is further decreased, the splitting of these two resonances increases [Figs.~\ref{fig:5}(c) and \ref{fig:5}(d)]. 
As depicted in Figs.~\ref{fig:5}(i)--\ref{fig:5}(l), this behavior can be understood by the characteristic behavior of the SW resonance with a local maximum in the AFM configuration at $|\mu_0 H| \approx \SI{60}{mT}$ and a decrease of the SAW excitation frequency.
For excitation frequencies $f$ below \SI{\sim 3}{GHz} and above \SI{\sim 7}{GHz}, the SW and SAW resonance frequencies do not match in the narrow field range available of the AFM configuration. 
There exists a window of excitation frequencies in which resonant SAW-SW interaction is possible in the AFM configuration of SAFs.
The range and position of this window can be modified by changing the properties of the SAF (e.g., $d^l, J_\text{eff}, M_s^l$).

In line with Fig.~\ref{fig:5}(i), resonant magnetoacoustic interaction at \SI{6.77}{GHz} is possible in the FM configuration at $|\mu_0 H| \approx \SI{500}{mT}$ and in the C configuration at $|\mu_0 H| \approx \SI{420}{mT}, \SI{200}{mT}$. 
%
%
The slope of $f(\mu_0 H)$ around the SAW-SW resonance fields and the SW linewidth $\Delta f_\text{HWHM}$ results in a narrow linewidth $\Delta H$ of the SAW-SW resonances in the FM and C configuration and in a broad resonance in the AFM configuration. 
%
%


In agreement with the discussion of Fig.~\ref{fig:4}, the magnetoacoustic SW excitation efficiency is governed by the static magnetization configuration with large, medium, and low efficiency in the AFM, C, and FM configuration, respectively.
Moreover, in the AFM configuration, the in-plane magnetoacoustic driving fields $\tilde{h}_2^l$ and $\Delta S_{21}$ are at maximum at $\phi_H \approx \SI{\pm 45}{\degree}$. 
In the C configuration, the symmetry of the magnetoacoustic driving field is complex, because the static magnetization directions $\phi_0^l$ are not parallel or antiparallel to the external magnetic field direction $\phi_H$. 
The dominating in-plane magnetoacoustic driving fields $\tilde{h}_2^l$, which are shown in Fig.~\ref{fig:4}, are for \SI{6.77}{GHz} at the positions of the SAW-SW resonances $|\mu_0 H| \approx \num{200}, \SI{420}{mT}$ in the C configuration at maximum at $\phi_H \approx \SI{0}{\degree}, \SI{\pm 90}{\degree}$.
Thus, the resonances of the SAW transmission $\Delta S_{21}$ in the C configuration in Figs.~\ref{fig:5}(a) and \ref{fig:5}(e) are also at maximum at $\phi_H \approx \SI{0}{\degree}, \SI{\pm 90}{\degree}$.
Note that the parameters of the SAF have a strong influence on the symmetry dependence of $\Delta S_{ij}$ in the C configuration.

Furthermore, the nonreciprocity of the SW dispersion causes a broken symmetry $f(+\mu_0 H) \neq f(-\mu_0 H)$ in Figs.~\ref{fig:5}(i)--\ref{fig:5}(l). According to the effective fields caused by iDMI and IDC in Eqs.~\eqref{eq:HeffDMI} and \eqref{eq:HeffIDC}, the frequency nonreciprocity $\Delta f_\pm$ decreases with a decrease of the wave vector $k$ and SAW excitation frequency $f$.  
Therefore, also the maximum nonreciprocity of the SAW transmission magnitude $\Delta S_\pm(\mu_0 H, \phi_H)=|\Delta S_{21}(\mu_0 H, \phi_H)-\Delta S_{12}(\mu_0 H, \phi_H)|$ decreases from \SI{\sim 2}{dB} at \SI{6.77}{GHz} to \SI{\sim 0.4}{dB} at \SI{3}{GHz}.

\subsection{Nonreciprocity}

\begin{figure*}
\includegraphics[scale=1]{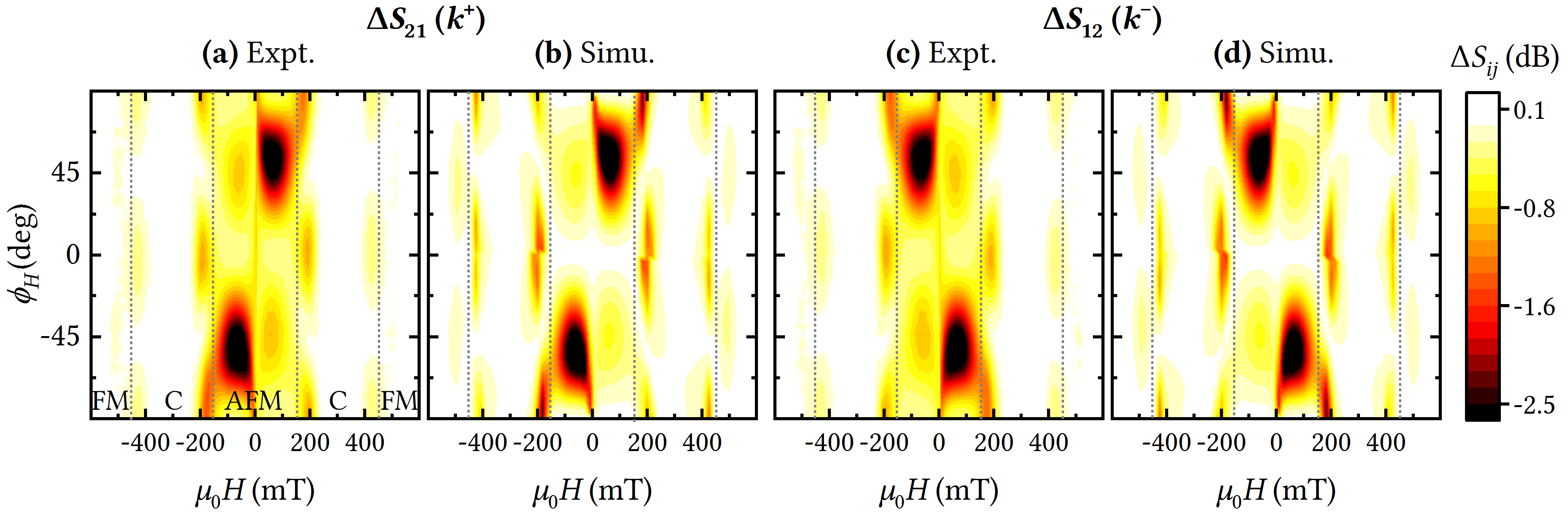}
\caption{
(a),(c)
Change of the SAW transmission $\Delta S_{ij}(\mu_0 H,\phi_H)$ for counter-propagating magnetoacoustic waves with the wave vectors $k^+$ and $k^-$ at \SI{6.77}{GHz}. 
The large frequency nonreciprocity $\Delta f_\pm$ of about \SI{2}{GHz} at $\phi_H=\pm \SI{45}{\degree}$ in the AFM configuration (see Fig.~\ref{fig:3}) results in a pronounced nonreciprocity of the SAW transmission.
(a),(c)
Experiment (Expt.) and (b),(d) corresponding simulations (Simu.) show good overall agreement.
\label{fig:6}
}
\end{figure*}




After having discussed the frequency dependence, we now look in more detail at the nonreciprocity for the frequency with the largest SAW-SW interaction at $f=\SI{6.77}{GHz}$. To this end, we show in Fig.~\ref{fig:6} the experimental and simulation results of the magnetoacoustic transmission $\Delta S_{21}(\mu_0 H, \phi_H)$ and $\Delta S_{12}(\mu_0 H, \phi_H)$ for counter-propagating magnetoacoustic waves with the wave vectors $k^+$ and $k^-$.
%

%
%

We observe an excellent agreement between experiment and simulation for both, $\Delta S_{21}$ and $\Delta S_{12}$. 
A pronounced nonreciprocity of the transmission $\Delta S_\pm$ shows only up in the AFM configuration.
As discussed next, the reason for the nonreciprocal transmission $\Delta S_\pm$ lays in the nonreciprocity of the SW dispersion. 


The large frequency nonreciprocity $\Delta f_\pm$ of the SW resonance in Fig.~\ref{fig:3}(b) is caused by the individual additive contributions of nonreciprocity mediated by iDMI and IDC.
Thus, in the AFM configuration for $\mu_0 H > 0$ and $\phi_H > 0$, resonant magnetoacoustic interaction is only possible for the SWs with $k^+$ and the change of the SAW transmission $\Delta S_{21}$ presented in Fig.~\ref{fig:6} is large. %
For counter-propagating SAWs ($k^-$), the transmission $\Delta S_{12}(\mu_0 H > 0, \phi_H > 0)$ shows a maximum with very low amplitude. 
%
We attribute this maximum to off-resonant magnetoacoustic interaction which is possible because of the large linewidth $\Delta f_\text{HWHM}$ of the SW resonance [see Fig.~\ref{fig:5}(i)].
Besides Gilbert damping, the large effective SW damping $\alpha^l$ results from a large inhomogeneous line broadening and the spin pumping effect~\cite{Tserkovnyak.2002, Ku.2020}. These two effects are very relevant due to the low thicknesses $d^l$ and the Ru- and Pt-interfaces of the magnetic layers.
While frequency nonreciprocity is very large with $\Delta f_\pm \approx \SI{3}{GHz}$ for the Damon-Eshbach geometry ($\phi_0^l=\SI{\pm90}{\degree}$), the maximum nonreciprocity of the SAW transmission magnitude is only moderate with $\Delta S_\pm \approx \SI{2}{dB}$. 
This is again attributed to the large effective SW damping $\alpha^l$.


In addition to the considered in-plane magnetoacoustic driving fields $\tilde{h}_2^l$, there are the smaller ($b_2^l \tilde{a}_{xz} < b_1^l \tilde{a}_{xx}$) out-of-plane magnetoacoustic driving fields $\tilde{h}_1^l$ [see Eq.~\eqref{eq:drivingfields2}], which could cause nonreciprocal transmission because of a helicity mismatch between the driving fields and magnetizations precession~\cite{Dreher.2012, Sasaki.2017, Ku.2020, Ku.2021b}.
Since $\tilde{h}_1^A$ and $\tilde{h}_1^B$ are parallel (antiparallel) in the FM configuration (AFM configuration) and have both a phase shift of $\pm\pi/2$ with respect to $\tilde{h}_2^l$~\cite{Sasaki.2017, Ku.2020, Xu.2020}, the total impact of $\tilde{h}_1^l$ on the excitation of the optical SW modes is low. In the FM  and AFM configuration, the total impact of $\tilde{h}_1^l$ on the magnetization precession almost cancels because the top Co(4) layer is thicker but has a lower magnitude $\tilde{h}_1^B \propto b_2^B \tilde{a}_{xz} = \text{i}\SI{0.22}{T}$ than the bottom Co(2) layer. 
Identical simulations of $\Delta S_{ij}(\mu_0 H, \phi_H)$ but with $b_2^B \tilde{a}_{xz} \gg \text{i}\SI{0.22}{T}$ or $b_2^B \tilde{a}_{xz} \ll \text{i}\SI{0.22}{T}$ show a nonreciprocal excitation efficiency because of the SAW-SW helicity mismatch effect, as demonstrated in Appendix~\ref{appendix: helicity mismatch}. For the investigated sample, the impact of the SAW-SW helicity mismatch effect on the nonreciprocal transmission is rather small.

\section{Conclusions}

In conclusion, we have experimentally studied the SAW-SW interaction in SAFs. Our experiments are in good agreement with analytical model calculations, including interlayer exchange coupling, iDMI, and IDC. 
The model describes the experimental findings of the studied Pt/Co(2)/Ru/Co(4)/Pt SAF for multiple frequencies and different external magnetic field magnitudes, which can be used to switch the equilibrium magnetization directions in both Co layers from AFM to C to FM configuration.
While the magnetoacoustic excitation efficiency of SWs is maximized in the AFM configuration, it is medium in the C configuration and low in the FM configuration. 
This is a consequence of the excited optical SW mode in the SAF and the directions and magnitudes of the magnetoacoustic driving fields in both Co layers. 
Efficient SAW-driven SW excitation should be even possible in compensated SAFs with zero net magnetization. This is in contrast to the small excitation efficiency of the optical SW mode by Oersted fields employing antennas~\cite{Zhang.1994}.
%

Moreover, we demonstrate a large nonreciprocity $\Delta f_\pm$ of the SW dispersion because of the individual additive nonreciprocal contributions of iDMI and IDC. 
In the Damon-Eshbach geometry ($\phi_0^l=\SI{\pm90}{\degree}$) and AFM configuration, both contributions give rise to  $\Delta f_\pm \approx \SI{3}{GHz}$ for SWs with $f=\SI{6.77}{GHz}$ and $k=\SI{13.1}{\micro m^{-1}}$. 
The nonreciprocity $\Delta f_\pm$ switches the sign and is much lower in the FM configuration. 
The strength of the interlayer exchange interaction has a strong impact on the static magnetization configuration, which in turn governs the SW resonance frequency $f(\mu_0 H)$  and nonreciprocity $\Delta f_\pm$ by iDMI and IDC. This explains the dependence of the magnetoacoustic response $\Delta S_{ij}$ on the nonmagnetic spacer layer thickness $d_s$, as experimentally observed in Ref.~\cite{Matsumoto.2022}.
Because of the high SW damping $\alpha$ and the low magnetic film thicknesses used, the maximum nonreciprocity of the SAW transmission $\Delta S_\pm \approx \SI{2}{dB}$ is moderate in comparison to magnetic bilayers and SAFs with low-damping and thicker ferromagnetic layers ($\Delta S_\pm \approx \SI{30}{dB}$)~\cite{Shah.2020, Ku.2021b, Matsumoto.2022}.

\begin{acknowledgments}
This work is funded by the Deutsche Forschungsgemeinschaft (DFG, German Research Foundation) – project number 492421737.
\end{acknowledgments}

\appendix

\section{Equilibrium state of the magnetization}
\label{appendix:sec:equilibrium M}

The equilibrium magnetization direction $\phi_0^l$ is calculated by numerical minimization of the static free energy $E_S$ per surface area~\cite{Strijkers.2000, Franco.2020} in the $xyz$ coordinate system
\begin{align}
    E_S = \sum_{l=A,B}
    &
    d^l M_s^l
        \bigg(
               -\mu_0 \mathbf{H}\cdot\mathbf{m}^l + \frac{1}{2} \mu_0 (M_s^l - H_k^l) (m_z^l)^2
        \nonumber \\
        &
        -\frac{1}{2} \mu_0 H_\text{ani}^l (\mathbf{m}^l \cdot \mathbf{u}^l)^2
        \bigg)
            -J_\text{eff} \mathbf{m}^A\cdot\mathbf{m}^B,
    \label{appendix:eq:equilibrium M}
\end{align}
where $\mathbf{m}^l = \mathbf{M}^l/M_s^l$ are the unit vectors of the magnetizations $\mathbf{M}^l$.
We consider the Zeeman energy, demagnetization field, which is partly compensated by a surface anisotropy field $H_k$, an in-plane uniaxial anisotropy field $\mu_0 H_\text{ani}$ with the easy-axis direction along $\mathbf{u}$ and bilinear interlayer exchange coupling.
The unit vector $\mathbf{u}$ is assumed to be orientated in the plane of the magnetic film and makes the angle $\phi_\text{ani}^l$ with the $x$ axis.
We neglect interlayer demagnetization fields since the lateral size of the fabricated sample is much larger than its thickness~\cite{Franco.2020_intensity, Franco.2020}.

\section{Effective dipolar fields}
\label{appendix:dipolar fields}

The effective fields caused by iDMI are taken from Moon et al.~\cite{Moon.2013}
\begin{equation}
    {\bf H}_\text{eff,123,DMI}^l = i \frac{2 D_\text{eff}^l}{\mu_0 M_s^l} k \sin(\phi_0^l)
    \begin{pmatrix}
        -m_2^l \\ m_1^l
    \end{pmatrix}.
    \label{eq:HeffDMI}
\end{equation}
%
%
Moreover, the effective fields of the bilinear interlayer exchange coupling
\begin{align}
    &
    \mathbf{H}_\text{eff,123,IEC}^A = \frac{J_\text{bl}}{d^A \mu_0 M_s^A}
    \begin{pmatrix}
        m_1^B \\ \cos(\phi_0^A - \phi_0^B) m_2^B \\ \cos(\phi_0^A - \phi_0^B)
    \end{pmatrix}
    \label{eq:HeffIEC}
\end{align}
and interlayer dipolar fields 
\begin{align}
    {\bf H}_\text{eff,123,IDC}^A & = -M_s^B G_0^A G_0^B \frac{d^B}{2} \text{e}^{-|k|d_s} \nonumber \\
    & \times
    \begin{pmatrix}
    -|k| m_1^B - i \sin{(\phi_0^B)} k \, m_2^B \\
    \sin{(\phi_0^A)} \sin{(\phi_0^B)} |k| m_2^B - i \sin{(\phi_0^A)} k \, m_1^B \\
    0
    \end{pmatrix},
\label{eq:HeffIDC}
\end{align}
are taken from Gallardo et al.~\cite{R.A.Gallardo.2019} and rewritten in the ($\mathbf{e}_1^l,\mathbf{e}_2^l,\mathbf{e}_3^l$)-coordinate system. Hereby, $G_0^l = \left(1-\text{e}^{- |k| d^l}\right) /  \left(|k| d^l\right) $ are dipolar SW terms~\cite{Kalinikos.}. 
For the effective fields $\mathbf{H}_\text{eff,123,IEC}^B$ and $\mathbf{H}_\text{eff,123,IDC}^B$ of layer $B$, the replacements $A \rightarrow B$, $B \rightarrow A$, and $-i \rightarrow +i$ are carried out.

\section{Sample preparation} 
\label{appendix:sample preparation}


After preparation of the IDTs, rectangular-shaped Pt(\num{3})/Co(\num{2})/Ru(\num{0.85})/Co(\num{4})/Pt(\num{3})/Si$_3$N$_4$(\num{3}) SAFs (thicknesses are given in \si{nm}) were deposited at room temperature by magnetron sputtering (base pressure \SI{<1E-8}{mbar}) between the IDTs. The Ar pressure was kept constant at \SI{3.5E-3}{mbar} during the deposition process and the sample holder was rotating during sputtering. The individual layer thicknesses were determined using a calibrated deposition rate (Pt: \SI{0.05}{nm/s}, Co: \SI{0.02}{nm/s}, Ru: \SI{0.025}{nm/s} and Si$_3$N$_4$: \SI{0.0065}{nm/s}). The Si$_3$N$_4$(\num{3}) capping layer prevents oxidation of the SAF.

\begin{figure*}
\includegraphics[scale=1]{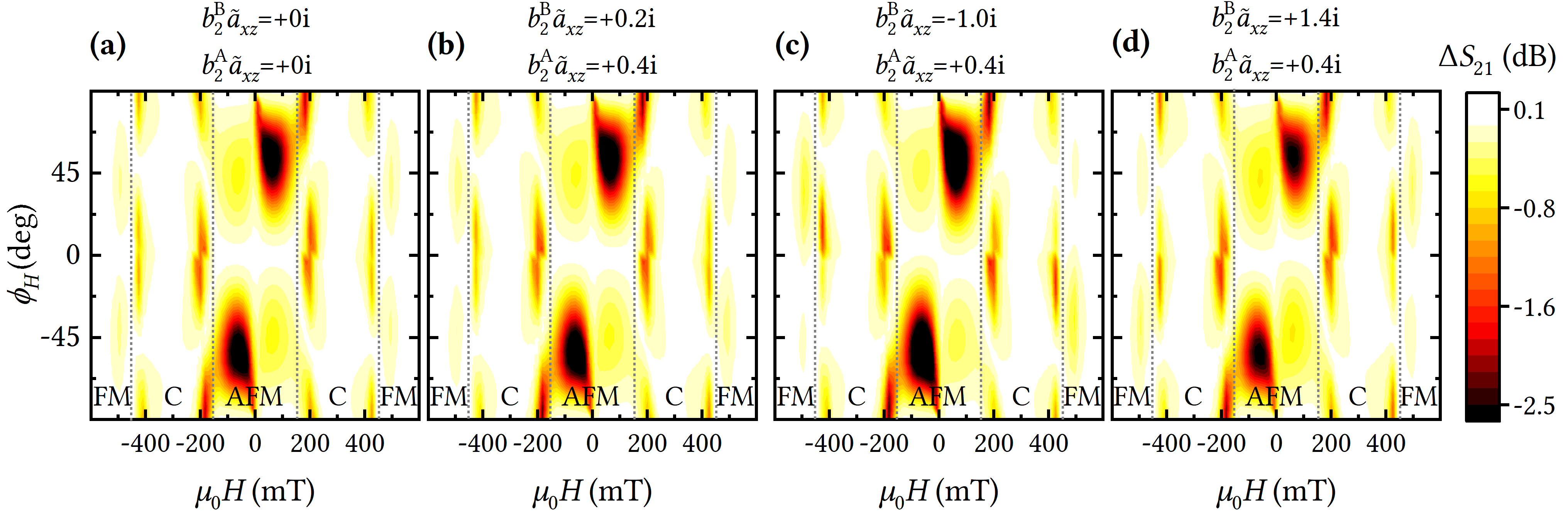}
\caption{
The out-of-plane magnetoacoustic driving field $\tilde{h}_1^l \propto b_2^l \tilde{a}_{xz}$ is changed in simulations to probe the impact of the SAW-SW helicity mismatch effect on the SAW transmission $\Delta S_{ij}(\mu_0 H,\phi_H)$ at \SI{6.77}{GHz}. 
(a)
The SAW-SW helicity mismatch effect vanishes for an in-plane linear polarized driving field. The results in (a) are almost identical to the results in (b), for which the same simulation parameters as in all previous simulations were used. Thus, the impact of the SAW-SW helicity mismatch effect is small in the simulations in Figs.~\ref{fig:5} and \ref{fig:6}.
If the driving field $\tilde{h}_1^B \propto b_2^B \tilde{a}_{xz}$ of the top Co layer is artificially low (c) or high (d), the SAW-SW helicity mismatch effect modulates slightly the SAW transmission nonreciprocity.
\label{fig:7}
}
\end{figure*}

\section{Simulation parameters} 
\label{appendix:simulation parameters}

In this section, we describe how the simulation parameters in Table~\ref{tab:1} were derived. 
First, we determined the effective interlayer coupling constant $J_\text{eff}$ from the $M$-$H$ hysteresis loop in Fig.~\ref{fig:2}(a).
%
We used the macrospin model to simulate  $M$-$H$. 
With the saturation magnetizations $M_s^l$ obtained by SQUID-VSM measurements performed on single layer reference samples and the nominal magnetic thin film thicknesses $d^l$, we obtain rather good agreement between experiment and simulation for $J_\text{eff}=\SI{-0.95}{mJ/m^2}$. 
This value is in the same range as reported in the literature for similar SAFs~\cite{Bloemen.1994, Strijkers.2000}. Discrepancies between experiment and simulation are attributed to the simplification of the macrospin approach~\cite{Tannous.2008}.
In agreement with the literature, biquadratic interlayer exchange coupling~\cite{Demokritov.1998, Rezende.1998, R.A.Gallardo.2019} is usually neglected to describe the magnetic behavior of Co/Ru/Co multilayers~\cite{Parkin.1990, Bloemen.1994, Strijkers.2000, R.Verba.2019, Kolesnikov.2018}.

Furthermore, for the simulations, we use $M_s^l$, the $g$-factor, $H_k^l$, and $\alpha^l$ obtained by SQUID-VSM and bbFMR measurements performed on the single layer reference samples. These parameters are summarized in Table~\ref{tab:1}. Hereby, the Gilbert damping $\alpha_\text{FMR}^l$ and the inhomogeneous line broadening $\Delta H_\text{FMR}^l$ contribute to the effective SW damping constant~\cite{Ku.2020}
\begin{equation}
    \alpha^l= \frac{\gamma^l}{2\omega} \mu_0 \Delta H_\text{FMR}^l + \alpha_\text{FMR}^l,
    \label{eq:alphaEff}
\end{equation}
where $\gamma^l$ is the absolute value of the gyromagnetic ratio. 
The magnetic exchange constant $A_\text{ex}^l=\SI{31}{pJ/m}$ is taken from the literature~\cite{Bertotti.1998, Franco.2020}. In addition, $D_\text{eff}^A$ and $b_1^A \tilde{a}_{xx}$ are taken from magnetoacoustic transmission measurements, which were performed on a LiNbO$_3$/Pt(\num{3})/Co(\num{2})/Ru(\num{0.85})/Si$_3$N$_4$(\num{3}) sample and carried out as described in Ref.~\cite{Ku.2020}.
Because of the interface character of iDMI and $d^A=0.5 d^B$, we assume that $D_\text{eff}^B = 0.5 D_\text{eff}^A$.

Finally, we use Eq.~\eqref{eq:5:S21Final} to simulate the experimental results of the SAF as presented in Figs.~\ref{fig:5} and \ref{fig:6}.
Experiment and simulation show good quantitative agreement under the assumptions for $b_1^B \tilde{a}_{xx}$, $b_2^B \tilde{a}_{xz}$, and the uniaxial anisotropies, given in Table~\ref{tab:1}.
Since the uniaxial anisotropies are small ($\mu_0 H_\text{ani}^l \leq \SI{4}{mT}$) in comparison to the external magnetic field magnitude ($|\mu_0 H| \leq \SI{600}{mT}$), the anisotropies have mainly an impact around zero field.
We attribute different magnetoelastic constants $b_{1,2}^A$ and $b_{1,2}^B$ of the bottom and top Co layer to different growth conditions caused by different seed layers and thicknesses.

\section{Impact of the helicity mismatch effect} 
\label{appendix: helicity mismatch}

The nonreciprocal SAW transmission in magnetoacoustic transmission measurements can be also caused by the nonreciprocity of the SW dispersion and/or the SAW-SW helicity mismatch effect. 
By changing the helicity of the magnetoacoustic driving field in the simulations, we additionally study the impact of the SAW-SW helicity mismatch effect on the SAW transmission. 
In Fig.~\ref{fig:7}, we show the simulated SAW transmission $\Delta S_{21}(\mu_0 H,\phi_H)$ with the parameters of Table~\ref{tab:1}, but with modified values for the out-of-plane magnetoacoustic driving field $\tilde{h}_1^l \propto b_2^l \tilde{a}_{xz}$ [see Eq.~\eqref{eq:drivingfields2}].
The SAW-SW helicity mismatch effect vanishes for $b_2^A \tilde{a}_{xz}=b_2^B \tilde{a}_{xz}=0\text{i}$ in Fig.~\ref{fig:7}(a), because the driving field is in-plane linear polarized. 
The results in Fig.~\ref{fig:7}(a) are almost identical to the results in Fig.~\ref{fig:7}(b), for which the same simulation parameters as in all previous simulations were used. Thus, the impact of the SAW-SW helicity mismatch effect on the nonreciprocal transmission in the previous simulations is very small.

If the driving field $|\tilde{h}_1^B|$ of the top Co layer is artificially large, the SAW-SW helicity mismatch effect modulates the SAW transmission nonreciprocity. 
For instance, with an increase of $b_2^B \tilde{a}_{xz}$ in Figs.~\ref{fig:7}(c), \ref{fig:7}(b), and \ref{fig:7}(d) the SAW absorption of the SAW-SW resonance in the AFM configuration increases in the 2nd quadrant and decreases in the 3rd quadrant. 
Additionally, the intensity of the resonances in the FM configuration differs between the 2nd and 3rd quadrant for the artificially large $|b_2^B \tilde{a}_{xz}|$ in Figs.~\ref{fig:7} (c) and \ref{fig:7} (d). 
Such an asymmetry of the intensity of the resonances in the FM configuration is not observed in the experimental results presented in Fig.~\ref{fig:5}(a). 
We conclude that the impact of the SAW-SW helicity mismatch effect on the nonreciprocal transmission of the investigated sample is rather small in comparison to the impact of the nonreciprocity of the SW dispersion.


%

\end{document}